\documentclass[utf8]{frontiersSCNS} 

\usepackage{url,hyperref,lineno,microtype,subcaption}

\usepackage{amsmath}

\usepackage{aas_macros}

\usepackage{color}
\usepackage{ulem}

\def\keyFont{\fontsize{8}{11}\helveticabold }
\def\firstAuthorLast{Marschall {et~al.}} 
\def\Authors{Raphael Marschall\,$^{1,*}$, Johannes Markkanen\,$^{2}$, 
Selina-Barbara Gerig\,$^{3,4}$, Olga Pinz\'on-Rodr\'iguez\,$^{3}$, Nicolas Thomas\,$^{3,4}$, and Jong-Shinn Wu\,$^{5}$}
\def\Address{$^{1}$Southwest Research Institute, Boulder Office, 1050 Walnut St Suite 300, Boulder, CO 80302, USA  \\
$^{2}$Max Planck Institute for Solar System Research, Justus-von-Liebig-Weg 3, 37077 Göttingen, Germany\\
$^{3}$Physikalisches Institut, Universit\"at Bern, Sidlerstr. 5, CH-3012 Bern, Switzerland\\
$^{4}$NCCR PlanetS, Sidlerstrasse 5, 3012 Bern, Switzerland 
$^{5}$Department of Mechanical Engineering, National Chiao Tung University, Hsinchu, Taiwan
}
\def\corrAuthor{Raphael Marschall}
\def\corrEmail{marschall@boulder.swri.edu}

\begin{document}
\onecolumn
\firstpage{1}

\title[The dust-to-gas ratio, size distribution, and dust fall-back fraction of comet 67P]{The dust-to-gas ratio, size distribution, and dust fall-back fraction of comet 67P/Churyumov-Gerasimenko: Inferences from linking the optical and dynamical properties of the inner comae.}

\author[\firstAuthorLast ]{\Authors} 
\address{} 
\correspondance{} 

\extraAuth{}

\maketitle

\begin{abstract}

\section{}
In this work, we present results that simultaneously constrain the dust size distribution, dust-to-gas ratio, fraction of dust re-deposition, and total mass production rates for comet 67P/Churyumov-Gerasimenko. We use a 3D Direct Simulation Monte Carlo (DSMC) gas dynamics code to simulate the inner gas coma of the comet for the duration of the Rosetta mission. The gas model is constrained by ROSINA/COPS data. Further, we simulate for different epochs the inner dust coma using a 3D dust dynamics code including gas drag and the nucleus' gravity. Using advanced dust scattering properties these results are used to produce synthetic images that can be compared to the OSIRIS data set. These simulations allow us to constrain the properties of the dust coma and the total gas and dust production rates.\\
We determined a total volatile mass loss of $(6.1 \pm 1.5) \cdot 10^9$~kg during the 2015 apparition. Further, we found that power-laws with $q=3.7^{+0.57}_{-0.078}$ are consistent with the data. This results in a total of $5.1^{+6.0}_{-4.9}\cdot10^9$~kg of dust being ejected from the nucleus surface, of which $4.4^{+4.9}_{-4.2}\cdot10^9$~kg escape to space and $6.8^{+11}_{-6.8}\cdot10^8$~kg (or an equivalent of $14^{+22}_{-14}$~cm over the smooth regions) is re-deposited on the surface. This leads to a dust-to-gas ratio of $0.73^{+1.3}_{-0.70}$ for the escaping material and $0.84^{+1.6}_{-0.81}$ for the ejected material. We have further found that the smallest dust size must be strictly smaller than $\sim30\mu$m and nominally even smaller than $\sim12\mu$m.

\tiny
 \keyFont{ \section{Keywords:} comets, coma, 67P/Churyumov-Gerasimenko, dust-to-gas ratio, size distribution, modelling, dust dynamics} 
\end{abstract}

\section{Introduction}
The European Space Agency's (ESA) Rosetta mission escorted comet 67P/Churyumov-Gerasimenko (hereafter 67P) from August 2014 to September 2016 along its orbit through the inner Solar System. It watched as the comet's activity started to develop at large heliocentric distances, come to its culmination at perihelion, and decline as the comet travelled out towards Jupiter's orbit. This long-term continuous monitoring of the comet's activity has provided an unprecedented wealth of data on this comet and its activity.\\

The observations revealed a complex bi-lobate shape \citep{Sierks2015,Preusker2017} and diverse morphology \citep{Thomas2015a}. As a comet approaches the Sun it is heated and the ices start sublimating and ripping with them dust particles. Thus one of the important questions to be answered was what the bulk of the comet was made of i.e. what the bulk refractory-to-volatile ratio is. In the simplified view where any ejected material is lost to space two measurements are sufficient to determine this ratio. First, the total mass loss during one apparition measured by the Radio Science Investigation (RSI) \citep{Paetzold2019}. Second, the total volatile mass loss which can be indirectly determined by the in-situ measurements of the gas density \citep[e.g.][]{Fougere2016,Laeuter2018,Combi2020} or remote sensing data \citep[e.g.][]{Migliorini2016,BockeleeMorvan2016,Marshall2017,Biver2019}. In this simple case, the refractory-to-volatile ratio can be immediately inferred from those two measurements. But the complex surface morphology has revealed large dust deposits \citep{Thomas2018} that indicate that possibly a large fraction of the ejected dust is re-deposited \citep{Thomas2015b}. If that is indeed the case, then the two above mentioned quantities cannot constrain the total dust mass ejected but rather only the dust mass escaping the nucleus gravity. Further, the process of dust fall-back obscures the emitted dust-to-gas ratio.\\

One way of constraining the amount of fall-back material would be to attempt to measure the actual change in elevation of the surface as a function of time from local or global digital terrain models (DTM). We cannot assess at this point if that is indeed feasible with the Optical, Spectroscopic and Infrared Remote Imaging System \citep[OSIRIS, ][]{Keller2007} data set. Another way is to couple the scattering properties of the dust with a dynamical model of the dust coma constrained by the brightness of the dust coma. In this work, we have adopted the latter approach and modelled the inner gas and dust comae for the entire Rosetta mission. We use Rosetta Orbiter Spectrometer for Ion and Neutral Analysis \citep[ROSINA, ][]{Balsiger2007} data to constrain the gas production rate and OSIRIS data for our dust models. To constrain the dust models we compare the dust coma brightness as measured by OSIRIS to synthetic model images. This process links several dust parameters that are otherwise not easily combined. In particular, we will show how the dust size distribution, the dust-to-gas ratio, the fraction of fall-back and the optical properties are inter-dependant and thus cannot be determined independently.\\

In Sec.~\ref{sec:method} we will describe the method used and lay out the assumptions we have made. Furthermore, we will point out the free parameters of the models, that need constraining through Rosetta data. Some theoretical considerations are presented in Sec.~\ref{sec:analytic-solution}. We will discuss the results of our work in Sec.~\ref{sec:results} and summarise and conclude our work in Sec.~\ref{sec:summary-conclusion}.

\section{Method}\label{sec:method}

In this work, we have used the modelling approach (and in particular our \texttt{DRAG3D} model for the dust coma) described in detail in \cite{Marschall2016}. This approach has been successfully applied for the analysis and interpretation of multiple Rosetta instruments, in particular ROSINA, MIRO (Microwave Instrument for the Rosetta Orbiter), VIRTIS (Visible and Infrared Thermal Imaging Spectrometer), and OSIRIS \citep{Marschall2016,Marschall2017,Gerig2018,Marschall2019}. While in previous work we have applied this approach to specific epochs of the Rosetta mission, we have employed it here to cover the entire mission period to study longer-term processes.\\
In the following, we will briefly repeat some of the most important parts of the modelling elements and refer to \cite{Marschall2016} for a detailed description. \\\\

\subsection{General assumptions}
\begin{figure}[h!]
	\includegraphics[width=\textwidth]{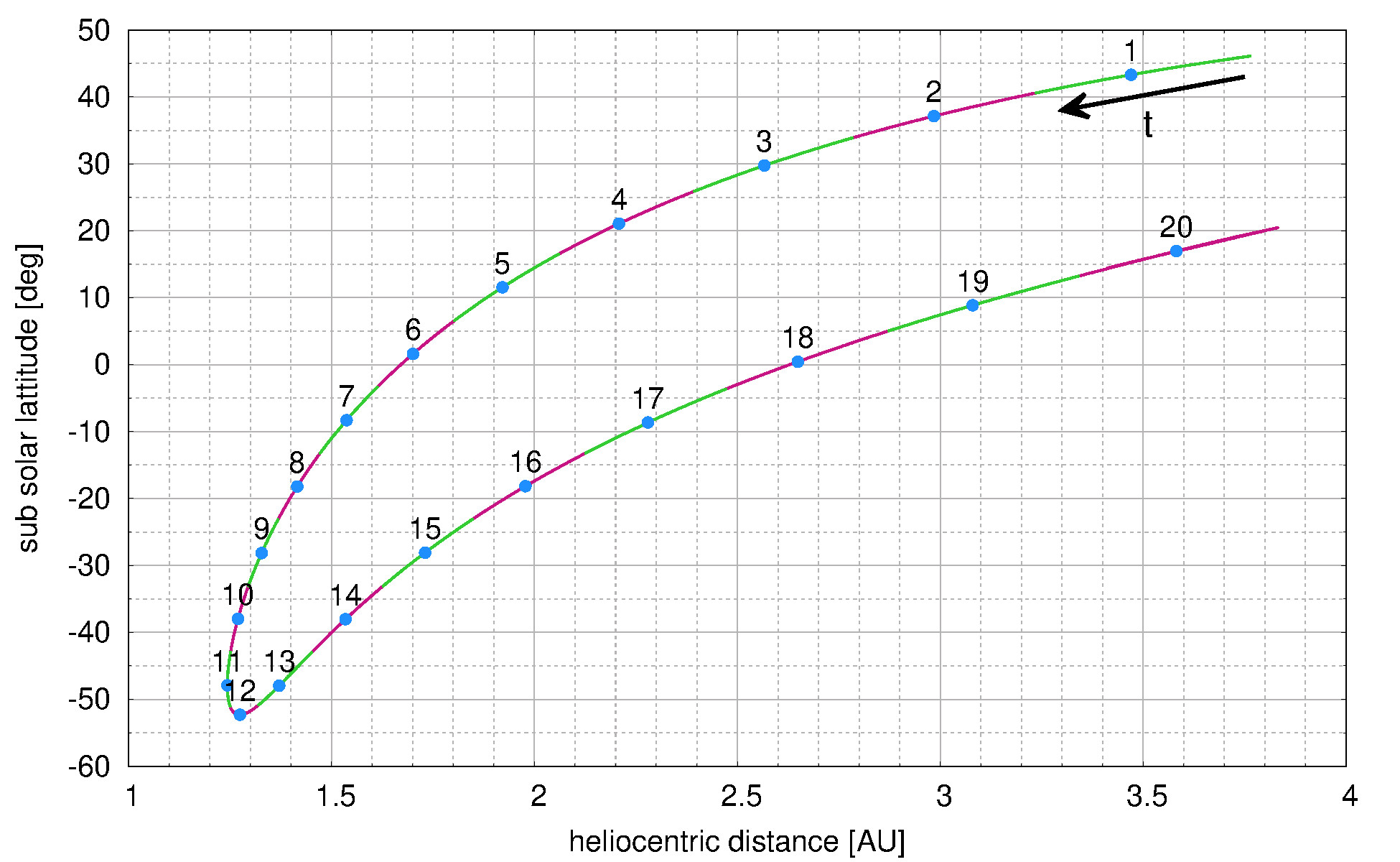}
	\caption{Heliocentric distance vs. sub-solar latitude of comet 67P during the escort phase of Rosetta as well as epochs used in this work.}
	\label{fig:epochs} 
\end{figure}

The calculation of the 3D gas flow field using the Direct Simulation Monte Carlo (DSMC) method is very computationally expensive and it is therefore currently not feasible to cover the entire escort phase of Rosetta (from August 2014 to September 2016) with a high temporal resolution. It is thus necessary to split the comet's orbit into a number of epochs that are computationally feasible and then interpolate between the results using a linear scaling between epochs. To ensure that the calculated results are representative of the respective epoch we make sure that during each of the epochs neither the total solar energy reaching the surface nor where the energy strikes the surface changes substantially. The amount of energy deposited is driven primarily by the heliocentric distance, $R_h$, while the location of deposition apart from the rotation of the comet is controlled by the sub-solar latitude, $LAT$. We thus chose that the inverse square of the heliocentric distance of the comet's location at the start and end time of each epoch shall be within $15\%$ of the location at the centre date of each epoch. Furthermore that the difference in sub-solar latitude be less than $5^{\circ}$ from the centre time of epoch to the start and end of the epoch, respectively. This leads to the twenty epochs listed in Tab.~\ref{tab:epochs} and illustrated in Fig.~\ref{fig:epochs}. Simulations were run for the centre time of each epoch. This choice also ensures that we cover the exact dates of the in- and outbound equinox (epochs 6 and 18) as well as perihelion (epoch 11) and summer solstice (epoch 12).\\

\begin{table}[] 
\caption{Start, centre, and end time for each of the epochs as well as the heliocentric distance and sub-solar latitude of the centre time of each epoch. All times are given in the format YYYY-MM-DD hhH UTC.\\}
\begin{tabular}{r|llllrl}\label{tab:epochs}
epoch & start time & centre time         & end time            & $R_h$ {[}AU{]} & $LAT$ {[}$^\circ${]} & qualifier        \\\hline \hline
1     & 2014-07-05 18H & 2014-08-26 12H & 2014-10-03 18H & 3.4703  & 43.3 &                  \\
2     & 2014-10-03 18H & 2014-11-11 12H & 2014-12-10 18H & 2.9844  & 37.1 &                  \\
3     & 2014-12-10 18H & 2015-01-10 00H & 2015-02-02 06H & 2.5667  & 29.8 &                  \\
4     & 2015-02-02 06H & 2015-02-26 18H & 2015-03-18 00H & 2.2083  & 21.1 &                  \\
5     & 2015-03-18 00H & 2015-04-05 00H & 2015-04-20 12H & 1.9213  & 11.6 &                  \\
6     & 2015-04-20 12H & 2015-05-04 06H & 2015-05-16 12H & 1.7006  &  1.6 & inbound equinox  \\
7     & 2015-05-16 12H & 2015-05-27 12H & 2015-06-06 18H & 1.5372  & -8.3 &                  \\
8     & 2015-06-06 18H & 2015-06-16 06H & 2015-06-25 06H & 1.4156  & -18.2    &                  \\
9     & 2015-06-25 06H & 2015-07-04 00H & 2015-07-12 12H & 1.3278  & -28.1    &                  \\
10    & 2015-07-12 12H & 2015-07-21 06H & 2015-07-30 18H & 1.2695  & -38.0    &                  \\
11    & 2015-07-30 18H & 2015-08-11 00H & 2015-08-21 12H & 1.2432  & -47.9    & perihelion       \\
12    & 2015-08-21 12H & 2015-09-02 18H & 2015-09-16 12H & 1.2742  & -52.3    & summer solstice  \\
13    & 2015-09-16 12H & 2015-09-27 12H & 2015-10-12 00H & 1.3709  & -48.0    &                  \\
14    & 2015-10-12 00H & 2015-10-25 06H & 2015-11-07 18H & 1.5344  & -38.0    &                  \\
15    & 2015-11-07 18H & 2015-11-22 00H & 2015-12-07 12H & 1.7307  & -28.1    &                  \\
16    & 2015-12-07 12H & 2015-12-24 12H & 2016-01-12 18H & 1.9778  & -18.2    &                  \\
17    & 2016-01-12 18H & 2016-02-01 18H & 2016-02-27 06H & 2.2796  & -8.6 &                  \\
18    & 2016-02-27 06H & 2016-03-22 12H & 2016-04-23 06H & 2.6491  & 0.4  & outbound equinox \\
19    & 2016-04-23 06H & 2016-05-24 00H & 2016-07-04 00H & 3.0797  & 8.9  &                  \\
20    & 2016-07-04 00H & 2016-08-13 18H & 2016-09-28 00H & 3.5819  & 17.0 &                 
\end{tabular}
\end{table}

The basis of all simulations is the 3D shape model by \cite{Preusker2017}. We use a decimated model with $\sim440'000$ facets due to our computational constraints. To fully define the illumination condition we need to select the sub-solar longitude in addition to the heliocentric distance and sub-solar latitude which are set by the choice of epoch. For each epoch we have run simulations for sub-solar longitudes of $0^\circ$, $30^\circ$, $60^\circ$, $90^\circ$, $120^\circ$, $150^\circ$, $180^\circ$, $210^\circ$, $240^\circ$, $270^\circ$, $300^\circ$, and $330^\circ$. This results in a total of 240 different illumination conditions for the entire mission period.\\

For each illumination condition, we calculate the incidence angle (angle between the surface normal and the direction of the Sun) of each facet taking into account self-shadowing. This allows calculating the solar energy entering the surface neglecting re-radiation from other facets. By means of a simple energy balance of the incoming solar energy, thermal re-radiation and sublimation we can calculate the sublimation temperature and the sublimation rate of each facet assuming pure water ice. 

We do not take into account any emission from shadowed facets, be it due to local night or mutual shadowing by other parts of the nucleus. The calculated pure ice sublimation rate of each facet needs to be scaled to match observed sublimation rates at 67P. Here we assume a pure H$_2$O ice surface that is areally mixed with inert refractory surface akin to a chequerboard pattern. This surface fraction of the facet covered by ice, which is a priori not known, is a free parameter of the model. We refer to this scaling factor as the \textit{effective active fraction} (EAF). This factor only has a physical interpretation for a pure ice surface where it would represents the fraction of pure ice of an areally mixed surface needed for a specific sublimation flux. In general though it is not a physical parameter and should not be interpreted as such.

In the next steps, we calculate the gas and dust flow fields in three dimensions. We then perform a column integration along the line-of-sight through the dust coma for a specific viewing geometry of the OSIRIS NAC (narrow-angle) and WAC (wide-angle) cameras \citep{Keller2007} and convolve the dust column densities with the optical properties of the dust to arrive at absolute radiance values that can be compared with the OSIRIS images. One major assumption that goes into this approach is that there is no significant back-coupling from the dust to the gas allowing a sequential treatment of the two flows. For low dust-to-gas mass ratios, this is certainly justified \citep{Marschall2016} but will break down when a lot of dust is released. We will further discuss this limitation later on. \\

\subsection{Gas kinetic simulations}

The gas flow-field is calculated using the DSMC technique.  The code used is called \texttt{UltraSPARTS}\footnote{http://www.plasmati.tw} and is a commercialized derivative of the \texttt{PDSC++} code \citep{Su13} used in previous papers \citep[e.g][]{Marschall2016,Marschall2017}. \texttt{PDSC++} is a C++ based, parallel DSMC code which is capable of simulating 2D, 2D-axisymmetric, and 3D flow fields. The code has been developed over the past 15 years \citep{Wu03,Wu04,Wu05} and contains several important features including the implementation of 2D and 3D hybrid unstructured grids, a transient adaptive sub-cell method (TAS) for denser flows, and a variable time-step scheme (VTS). In the parallel version, computational tasks are distributed using the Message Passing Interface (MPI) protocol. The improved \texttt{UltraSPARTS} (Ultra-fast Statistical PARTicle Simulation Package) has been applied to 67P \cite{Gerig2018,Marschall2019}. Here we simulate the full 3D gas flow up to a distance of 10~km from the nucleus centre.\\

The sublimation temperature and flux -- calculated as described above -- for each facet are set as initial conditions of the simulation. This includes implicitly the assumption of the appropriate EAF. We assume here that the EAF of all facets are the same (i.e. homogeneous surface properties) but can change from epoch to epoch. This results in one value for the global EAF per epoch. Though we know from previous works \citep[e.g][]{Bieler2015,Marschall2016,Fougere2016,Zakharov2018a} that there are regional inhomogeneities that can be encoded in EAF it is not the focus of this work to constrain these inhomogeneities. Rather we seek a global estimate of the fluxes and dynamical behaviours. Because the EAF is a free parameter it needs to be constrained by data. In our case, we determine the EAF by comparing modelled densities extrapolated to the Rosetta position and actual COmet Pressure Sensor (COPS; \cite{Balsiger2007}) measurements during each epoch. Within each epoch where we match the sub-solar longitude of a measurement, we extract from the respective simulation the gas number density at the position of the spacecraft. If the spacecraft distance is larger than $10$~km we extrapolate the value from the $10$~km surface to the spacecraft distance assuming free radial outflow. This assumption is well justified as shown in \cite{Marschall2016}. Though this does not capture the detailed structure of the ROSINA/COPS data it does account accurately for the average activity level at each epoch. Tab.~\ref{tab:EAF-and-Q} shows the EAF used and the resulting average global H$_2$O production rate for one comet day.

\begin{figure}[h!]
	\includegraphics[width=\textwidth]{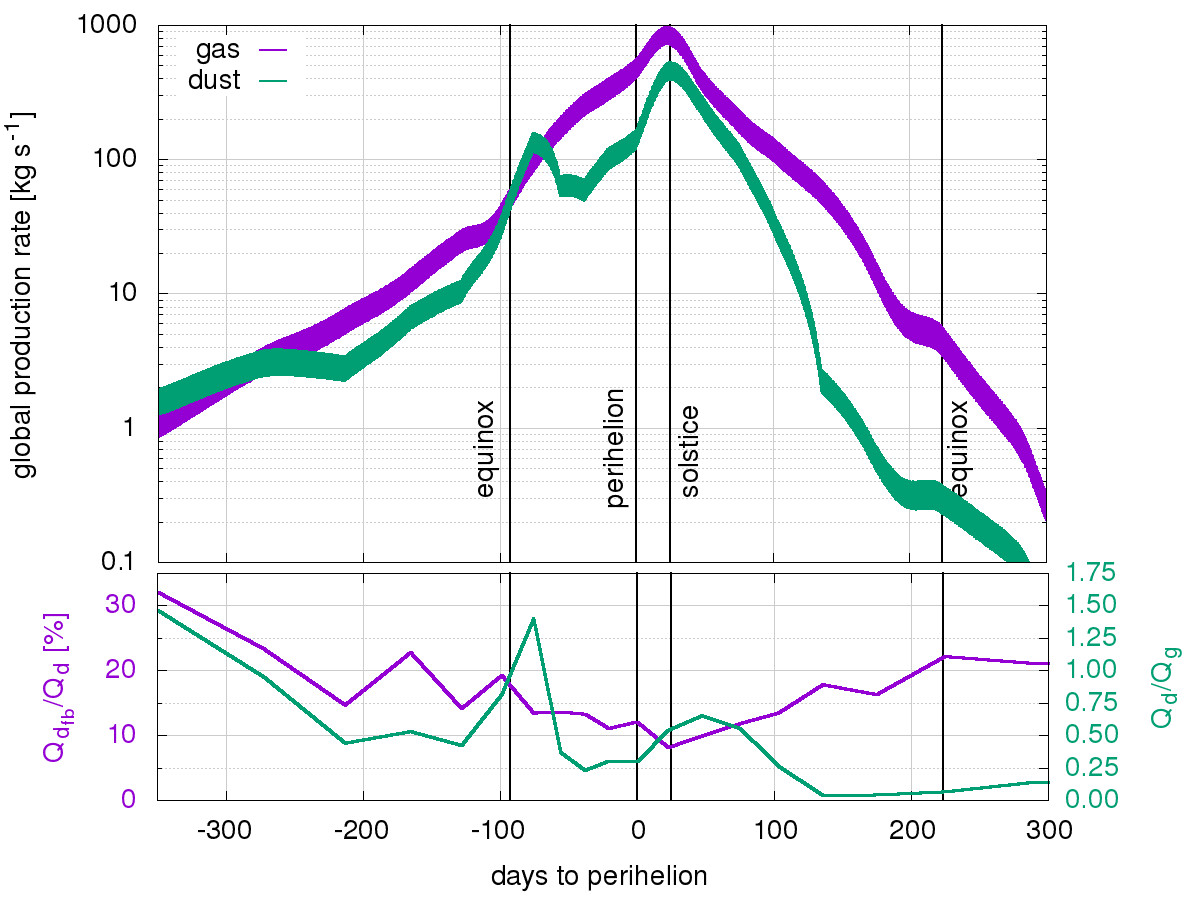}
	\caption{The top panel shows the global production rate for the gas (purple) and dust (green) as a function of days to perihelion. The bands indicate the range due to the diurnal variation. The gas production rates have been constrained by ROSINA/COPS measurements while the dust production rates are from combined constraints of OSIRIS and gas fluxes. For the dust a minimum size of $r_{min} = 0.1~\mu$m and power-law exponent of $q=3.75$ are assumed. The bottom panel shows the fraction of dust fall-back (purple) and dust-to-gas ratio (green).}
	\label{fig:Q-vs-epoch} 
\end{figure}

\begin{table}[]
\caption{The effective active fraction (EAF) assumed and the resulting average H$_2$O production rate for each epoch.}
\begin{tabular}{r|ll}\label{tab:EAF-and-Q}
epoch & EAF    & Q$_{H_2O}$ [kg s$^{-1}$]       \\ \hline\hline
1     & 0.87  & 1.047   \\
2     & 1.40  & 3.120   \\
3     & 1.60  & 6.178   \\
4     & 2.10  & 12.42  \\
5     & 2.80  & 24.74  \\
6     & 3.24  & 38.54  \\
7     & 6.00  & 92.38  \\
8     & 8.80  & 168.2  \\
9     & 10.9  & 249.9 \\
10    & 12.5  & 328.9 \\
11    & 16.5  & 473.2 \\
12    & 30.1  & 827.0 \\
13    & 17.3  & 393.3 \\
14    & 11.4  & 192.7 \\
15    & 9.14  & 110.1 \\
16    & 6.84  & 55.98  \\
17    & 2.61  & 13.92  \\
18    & 1.33  & 4.321   \\
19    & 0.32  & 0.592   \\
20    & 0.41  & 0.393  
\end{tabular}
\end{table}
The global gas production rate as a function of time is shown in Fig.~\ref{fig:Q-vs-epoch} (purple band in top panel). Because we have used the total ROSINA/COPS data -- which also contains the other gas species other than water -- to constrain our emission, these values should be understood as a proxy for the entire emission. We have interpolated between the epochs using a local second-order polynomial. The fit for each epoch, i, includes three epochs i-1, i, i+1. The fitting parameters to calculate the mean gas production rate as a function of ephemeris time (ET) is shown in Tab.~\ref{tab:fitting-parameters-q}. The resulting integrated mass loss over the shown period adds up to $(6.1 \pm 1.5) \cdot 10^9$~kg. This is well in line with values published in other works as e.g. $(6.3 \pm 2.0) \cdot 10^9$~kg \citep{Combi2020} or $(5.8 \pm 1.8) \cdot 10^9$~kg \citep{Laeuter2018}. The error arises from the uncertainty of the data (up to 15\%; \cite{Tzou2017} although the relative errors are probably smaller (M. Rubin, pers. comm.)) and our model (5-10\%; \cite{Finklenburg14}) as well as from the scatter from the comparison of the data and our model. As it was not the goal of this work to constrain as precisely as possible the surface-emission distribution it is nevertheless noteworthy that our estimates come so close to the other published values. This illustrates that it is not necessary to know the surface-emission distribution well to estimate the total global volatile loss. Rather simple assumptions of the surface response is sufficient for such an estimate. Though it is not that surprising, because as pointed out in \cite{Marschall2019c} the global gas production rate can be fairly well estimated by even simplified models. Our peak production rate is reached at the summer solstice (epoch 12) and not perihelion (epoch 11) and therefore roughly 22~days post-perihelion. This is in line with dust coma measurements by OSIRIS. \cite{Gerig2018} reported peak dust coma brightness 20 days post-perihelion. This also hints at the fact that the obliquity plays an important role in the activity of comets. Though the heliocentric distance still is the main driver of the gas and dust activity ($O(1)$) it is the obliquity/season that controls the second order. The coincidence of the peak gas activity with the peak dust activity also indicates that the dust activity is mainly driven by H$_2$O or at least near-surface volatiles without a significant thermal lag.\\\\

\subsection{Dust dynamic simulations} \label{sec:dust-dynamics}

After the gas flow field has been evaluated, we calculate the dust flow field by injecting dust test particles into the flow. We use a typical approach for computing the dust motion in a gas flow-field taking into account gas drag and the comet gravity using our \texttt{DRAG3D} dust coma model detailed in \cite{Marschall2016} and the references therein. We assume that the dust mass production rate is proportional to the gas mass production rate and that the dust size distribution does not vary across the surface except in cases where certain dust sizes are no longer lifted because the gas pressure is too low to surpass the local gravity. The dust size distribution is thus only naturally modified by the dynamics and lifting process. It is assumed that the dust particles are at rest on the surface (i.e. the
ejection velocity is 0). The dust-to-gas mass ratio as well as the dust size distribution at the surface are free parameters of the model and will be constrained by the data as described below. Due to the presence of gravity, large dust particles may not reach escape speed and eventually return to the surface. The flux of back-fall particles is thus a further output of the model. It is important to note that we assume that the dust particles are desiccated, i.e. contain no significant amounts of volatiles that evaporate while airborne. They may still be wet but do not outgas significantly. This is a consequence of our assumption that there is no significant back-coupling of the dust flow onto the gas flow.\\

\begin{table}[]
\caption{List of OSIRIS images used in each of the epochs as well as their filter, central wavelength ($\lambda_c$), phase angle ($\alpha$), and cometo-centric distance (D$_{cc}$). The name of the OSIRIS image contains the camera used (WAC - wide angle camera, NAC - narrow angle camera) and the time stamp in the format \{YYYY-MM-DD\}T\{hh.mm.ss.ms\}.\\}
\begin{tabular}{r|lllll}\label{tab:osiris-images}
epoch & OSIRIS image name  & filter & $\lambda_c$ [nm] & $\alpha$ [$^\circ$] & D$_{cc}$ [km] \\ \hline\hline
1  & WAC\_2014-08-16T13.01.44.647 & F18 & 612.6 & 37.85 & 100.03 \\
2  & - & -   & - & -  & -   \\
3  & WAC\_2015-02-09T19.10.19.184 & F18 & 612.6 & 87.43 & 105.61 \\
4  & WAC\_2015-02-19T23.11.20.158 & F18 & 612.6 & 73.64 & 136.20 \\
5  & WAC\_2015-04-11T08.14.54.119 & F18 & 612.6 & 88.32 & 140.96 \\
6  & WAC\_2015-05-05T11.27.54.523 & F18 & 612.6 & 64.68 & 165.48 \\
7  & WAC\_2015-05-22T06.36.34.903 & F18 & 612.6 & 59.82 & 134.89 \\
8  & WAC\_2015-06-16T03.22.37.523 & F18 & 612.6 & 89.06 & 220.19 \\
9  & WAC\_2015-07-04T11.11.58.808 & F18 & 612.6 & 89.83 & 176.63 \\
10 & WAC\_2015-07-19T00.20.07.696 & F18 & 612.6 & 89.25 & 181.28 \\
11 & WAC\_2015-08-09T09.13.16.574 & F18 & 612.6 & 89.09 & 306.41 \\
12 & WAC\_2015-09-02T07.58.47.075 & F18 & 612.6 & 72.76 & 414.13 \\
13 & WAC\_2015-09-25T11.39.29.053 & F18 & 612.6 & 56.35 & 635.77 \\
14 & NAC\_2015-10-14T07.03.45.244 & F22 & 649.2 & 64.45 & 497.73 \\
15 & WAC\_2015-11-22T21.43.06.876 & F18 & 612.6 & 89.23 & 127.75 \\
16 & WAC\_2015-12-30T08.12.53.526 & F18 & 612.6 & 89.58 & 87.682 \\
17 & WAC\_2016-01-13T07.03.19.181 & F18 & 612.6 & 88.05 & 87.974 \\
18 & WAC\_2016-03-25T02.15.44.556 & F18 & 612.6 & 108.2 & 270.11 \\
19 & WAC\_2016-04-13T18.47.58.431 & F18 & 612.6 & 76.41 & 106.43 \\
20 & - & -   & -  & -  
\end{tabular}
\end{table}

For each epoch (except for two) we have selected one OSIRIS image where the illumination conditions of the image match one of the gas simulations (see Tab.~\ref{tab:epochs}). The images used in this work, as well as some camera and geometric properties, are shown in Tab.~\ref{tab:osiris-images}. Two main criteria were used to select these images. First, the images needed a large enough field of view such that projected distance (impact parameter, $b$) in the image plane from the centre of the comet to the edge of the image at each side was at least 9~km. Why this is an important constraint will be described in the next paragraph. Second, images need a sufficient signal to noise such that the dust coma brightness could be measured well. These two constraints unfortunately, eliminated all images for epoch 2 and 20. Epoch 2 included the 10~km orbit phase and thus did not provide large enough fields-of-view while the signal-to-noise was bad in epoch 20 due to the very low activity of the comet. Most images we have used were taken by the wide-angle camera (WAC) and filter 18 (central wavelength, $\lambda_c = 612$~nm) and at cometocentric distances between 87 and 635~km and phase angles between 37$^\circ$ and 108$^\circ$.\\

The dust field is calculated for each image using 41 different dust sizes from $10$~nm to 1~m. The dust sizes are logarithmically spaced with five dust sizes per decade. The particles are assumed to be spherical and have a density of $533$~kg~m$^{-3}$ matching roughly the bulk density of the nucleus \citep{Preusker2017}. Even though all dust sizes are simulated, not all of them contribute to the dust brightness in the coma. This is because the particles larger than a certain size might not all be lifted because the gas pressure cannot overcome gravity. Thus the number of dust sizes present in the coma depends on the heliocentric distance (epoch). The upper size limit (largest liftable size) is thus naturally determined and thus an outcome of the simulation. What the smallest dust size should be is unless. The smallest diameter of particle sub-units measured by MIDAS \citep{Mannel2019} is $100$~nm. Whether these could also be the smallest dust particles in the coma or if these measurements have an in-situ collection bias at the spacecraft is not clear. One could imagine that very small particles might not have been collected because of spacecraft and dust charging. \cite{DellaCorte2019} showed that particles, for which the ratio of the particle charge to its kinetic energy entering the electrostatic field of the space craft $q/E_k> 0.24 C J^{-1}$, will not reach the spacecraft. We will therefore leave this issue open for the moment and examine the impact of the smallest size on the results in Sec.~\ref{sec:results}.\\

Once the 3D dust field is simulated we calculate the dust column densities of each size for the specific viewing geometry of the respective OSIRIS image. The final image is composed by weighting the different dust sizes according to a specific dust size distribution and convolving the column densities with the scattering properties described in Sec.~\ref{sec:scattering}.

For each of the images we compare the integrated radiance of the dust coma along an aperture with impact parameter $b=11$~km and compare it to that of the synthetic images. Again it is not our goal in this work to match the structure of the inner dust coma but rather the overall global behaviour. \cite{Gerig2018} showed from OSIRIS data that the dust flow goes over to free radial outflow at an average impact parameter of $b\sim11$~km. This is in line with theoretical considerations of dusty flows \citep{Zakharov2018b}. Beyond that point, the dust brightness falls off with $1/b$ as expected for a freely expanding radial flow. For that reason, we have chosen $b=11$~km to be within the free-flow regime. If the field-of-view was not large enough we used the maximum available impact parameter.\\

\subsection{Scattering model} \label{sec:scattering}
\begin{figure}[h!]
	\includegraphics[width=\textwidth]{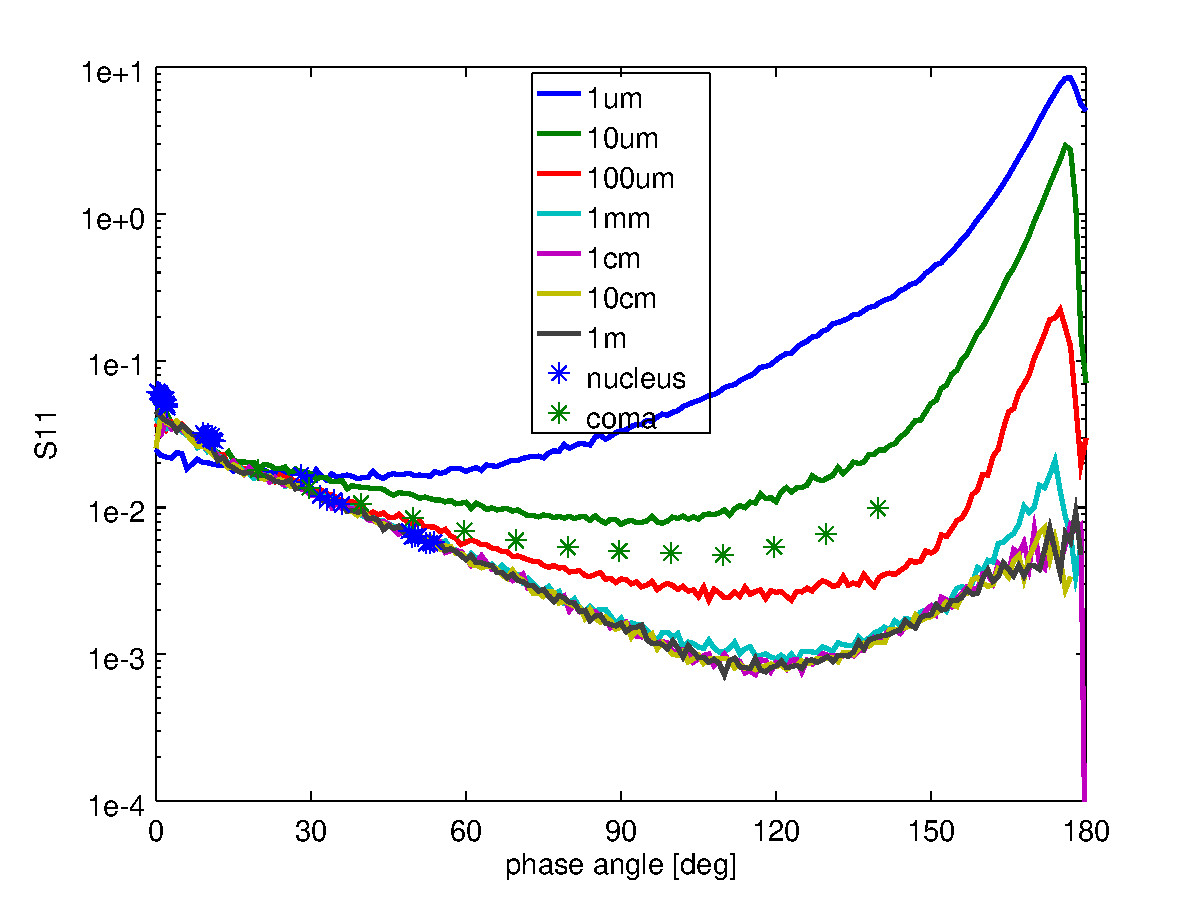}
	\caption{Shows the phase functions, S$_{11}$, as a function of phase angle for different particle sizes (solid lines). The green stars represent the measurements of the coma phase function by \cite{Bertini2017} while the blue stars show the nucleus phase function as measured by \cite{Fornasier2015,Feller2016,Masoumzadeh2017} }
	\label{fig:phase-functions} 
\end{figure}

Previously, we have used a spherical particle model and a Mie scattering code in our modelling pipeline. Here we use a much more sophisticated approach based on the recently introduced radiative transfer with reciprocal transactions framework \citep{Muinonen2018,Markkanen2018b}. The approach allows for scattering analysis of large irregularly shaped particles with wavelength-sized details. Here, the dust particles are considered to be irregular aggregates composed of sub-micrometre-sized organic grains and micrometre-sized silicate grains. Such a particle model has been found to be in good agreement with OSIRIS \citep{Markkanen2018} and VIRTIS \citep{Markkanen2019} phase function measurements.  The refractive index for silicate grains is assumed to be $m =  1.6048692 + i0.0015341$ corresponding to magnesium iron pyroxine \citep{Dorschner1995} and for organic grains $m = 1.55950 + i0.42964$ corresponding to amorphous carbon \citep{Jager1998}. At 612~nm (WAC filter 18) the resulting scattering phase functions for different particle sizes normalized to the geometric albedo are shown in Fig.~\ref{fig:phase-functions}. The figure shows good agreement of the phase function of large particles ($>1$~cm) with the nucleus phase function as measured by OSIRIS \citep{Fornasier2015,Feller2016,Masoumzadeh2017}. This should indeed be the case because larger particles should behave more and more like "small comets" themselves and thus be representative of the nucleus scattering properties. For small particles, the best agreement of a single dust size with the coma phases function \citep{Bertini2017} is between 10 and 100~$\mu$m. The numerical method of \citet{Markkanen2018} is not applicable to particles smaller than $1\mu$m. Thus for the particles smaller than $1\mu$m we use a Mie scattering code to determine the scattering properties \citep[see][]{Marschall2016} matching the single scattering albedo of the Mie result with the approach of \citet{Markkanen2018} for $1\mu$m particles. This gives us a smooth transition from the large particle region to the Rayleigh scattering region where particle's shape has a negligible effect on its scattering properties. This is a state of the art model and we have thus used its results throughout this work. But because the scattering model does have an effect on the results a re-evaluation of the results can be done if and when a better model arises.

\section{Theoretical consideration}\label{sec:analytic-solution}
To put some of our results in the next section (Sec.~\ref{sec:results}) into context, we present first some general theoretical considerations of what we can expect, in particular with regards to the relationship between the dust-to-gas ratio and the dust size distribution. We thus consider first a simple model where the comet is represented as a sonic (i.e. the gas velocity near the surface is equal to the local sound velocity and defined by the thermodynamic properties of the gas and the surface temperature i.e. $R_h$) spherical source of ideal perfect gas (with specific heat ratio $\gamma$=1.33) accelerating spherical solid grains. The source shall have radius $R_N$, nucleus mass $M_N$, total gas production rate $Q_g$ (kg/s).  The motion of a spherical grain in a flow from such a source was studied in \cite{Zakharov2018b} for a wide range of conditions. They defined

\begin{equation}\label{eq:Iv}
{\rm Iv} = \frac{3 Q_g}{32 R_N r \rho_d \pi v_{g0}}
,\end{equation}
and 
\begin{equation}\label{eq:Fu}
{\rm Fu} = \frac{G M_N}{R_N} \frac{1}{(v_g^{max})^2}
\end{equation}
which are dimensionless parameters, where $\rho_d$ is the specific density of the dust particles, $v_{g0}$ the gas velocity near the surface, $v_g^{max}$ the theoretical maximal velocity of gas expansion (defined by the thermodynamic properties of the gas and the surface temperature i.e. $R_h$), and $G$ the gravitational constant. $Iv$ characterises the ability of a dust particle to adjust to the gas velocity while $Fu$ quantifies the importance of gravity. \cite{Zakharov2018b}) found that for $Iv < 0.1$ (which is the case of 67P, and dust sizes $> 1$ nm) the dust particles reach 90\% of their terminal velocity at about $6\cdot R_N$. The terminal velocity of particles with radius, $r$, varies as $v_d(r) \propto r^{-0.5}$ for small $Fu$ (i.e. if gravity plays a minor role).
The asymptotic dust velocities are given by: 
\begin{equation}
 v_d(r) =\left(\frac{\rm{Iv}(r)}{\rm{Iv}(r_\ast)}\right)^{1/2} v_d(r_\ast) =
 \left(\frac{r}{r_\ast}\right)^{-1/2} v_d(r_\ast) = r^{-1/2} C_{Iv}
\label{eqn_vd}
\end{equation}
where $r_\ast$ and $v_d(r_\ast)$ are some referential size and corresponding terminal velocity, and $C_{Iv}$ is a constant.

For a dust size distribution given by a power-law, $r^{-q}$, the normalised mass distribution, $f_{md}$, of particles ejected from the surface is

\begin{equation}
f_{md}(r)=\left\{
 \begin{array}{ll}
  \frac{4-q}{r_{max}^{4-q}-r_{min}^{4-q}} r^{3-q}, & q\not=4 \\
  \ln{\left(-\frac{r_{max}}{r_{min}}\right)} r^{3-q}, & q=4
 \end{array}
 \right.
\label{eqn_fmd}
\end{equation}
where $r_{min}$ and $r_{max}$ are the smallest and largest dust sizes
ejected from the surface. In the following we will not considered specially the case of $q=4$. The dust production rate, $Q_d$, of each dust size is

\begin{equation}
Q_d(r)=\chi f_{md}(r) Q_g dr
,\end{equation}
where $\chi= Q_d/Q_g$ is the total dust-to-gas mass loss rate. Therefore, the number density of dust particles with radius, $r$, at the radial distance, $R$, from the centre of the nucleus is:

\begin{equation}
n(r,R)=\frac{\chi f_{md}(r)Q_g dr}{v_d(r) m_d(r)4 \pi R^2}
.\end{equation}

The column density at the distance $\varrho$ from the centre of the nucleus in the image plane is:

\begin{equation}
n_{col}(r,\varrho)= \int_{-\infty}^{+\infty} n(r,R) dz = \frac{\chi f_{md}(r)Q_g dr}{v_d(r) m_d(r)4} \frac{1}{\varrho}
\label{eqn_ncol}
.\end{equation}

The total number of dust particles in a column within a circular observing aperture of radius $\Re$ is:

\begin{equation}
N(r,\Re) =  \int_0^{\Re}n_{col}(r,\varrho) 2 \pi \varrho d\varrho = \frac{\chi f_{md}(r) Q_g dr}{v_d(r) m_d(r) 2} \pi \Re
\label{eqn_N}
.\end{equation}

The brightness is proportional to the flux $F$ ($W/m^2$) gathered by an instrument which for an optically thin coma is:

\begin{equation}
F(r,\Re)=\mathcal{F} N(r,\Re) \pi r^2 q_{sca}(r) \frac{\varphi_{av}(r)}{4 \pi} \frac{1}{\Delta^2 }
\label{eqn_flux_src}
\end{equation}
where $\mathcal{F}$ is the incident flux, $\Delta$ is observational distance, $q_{sca}$ is scattering efficiency and $\varphi_{av}$ is the phase function averaged over phase angle. Substituting Eqs.~(\ref{eqn_vd}), (\ref{eqn_fmd}) and (\ref{eqn_N}) in (\ref{eqn_flux_src}) we get:

\begin{equation}
F(r,\Re)=\frac{3}{32} \frac{\mathcal{F} Q_g \Re}{C_{Iv} \rho_d \Delta^2} \frac{4-q}{r_{max}^{4-q}-r_{min}^{4-q}} r^{\frac{5}{2}-q} \chi q_{sca}(r) \varphi_{av}(r) dr
\label{eqn_flux}
.\end{equation}

For fixed $\mathcal{F}$, $Q_g$, $R_N$, $\rho_d$, $v_{g0}$, $r_\ast$, $v_d(r_\ast)$, $\Delta$ and $\Re$
\begin{equation}
F(r,\Re)= C \frac{4-q}{r_{max}^{4-q}-r_{min}^{4-q}} \chi r^{\frac{5}{2}-q} q_{sca}(r) \varphi_{av}(r) dr
\label{eqn_I}
\end{equation}
where $C=\frac{3}{32} \frac{\mathcal{F} Q_g \Re}{C_{Iv} \rho_d \Delta^2}$ is a constant (for the given observational conditions).

\begin{figure}
 \centering
 \includegraphics[width=\textwidth]{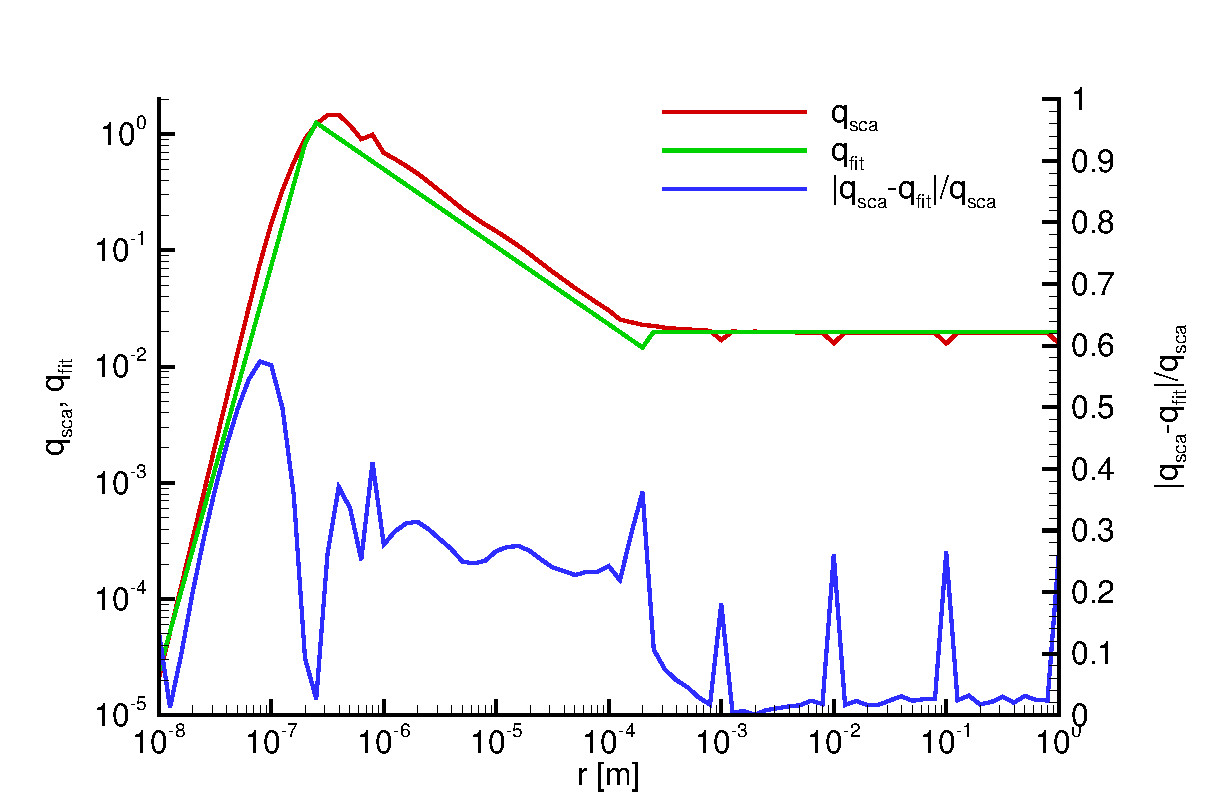}
 \caption{Scattering efficiency as a function of dust radius: computed (comp, red) according to Sec.~\ref{sec:scattering}, and fitted (fit, green). The relative difference between the computed and fitted is shown in blue.}		
 \label{fig_qfit2}
\end{figure}

For the optical properties, we make some simplifying assumption. First, we approximate the scattering efficiency $q_{sca}(r)$ to be:

\begin{equation}
   \small
   q_{sca}(r)=\left\{
   \begin{array}{ll}
         0.233 \cdot 10^{24} \cdot r^{7/2},     & 10^{-8} \le r<2
\cdot 10^{-7}  \\
         4.993 \cdot 10^{-5} \cdot r^{-2/3},     & 2 \cdot 10^{-7}\le
r < 2 \cdot 10^{-4} \\
         0.02,                                                 & 2
\cdot 10^{-4} \le r \le 1.0
   \end{array}
   \right.
   \normalfont
\label{eqn_qsca}
\end{equation}
Figure~\ref{fig_qfit2} shows the computed $q_{sca}$ from Sec.~\ref{sec:scattering}, the fitted $q_{fit}$ scattering efficiency and the relative difference. In this fit, we used ``round numbers’’ (i.e. this is a very rough fit). This simplification results in differences of $<50\%$. For the phase function $\varphi_{av}$ we assume it to be constant for all sizes. We estimate an error of the order of a factor of 2 from this simplification.

Under the assumption we made the integration of Eq.~(\ref{eqn_I}) becomes trivial.

\begin{figure}
 \centering
 \includegraphics[width=\textwidth]{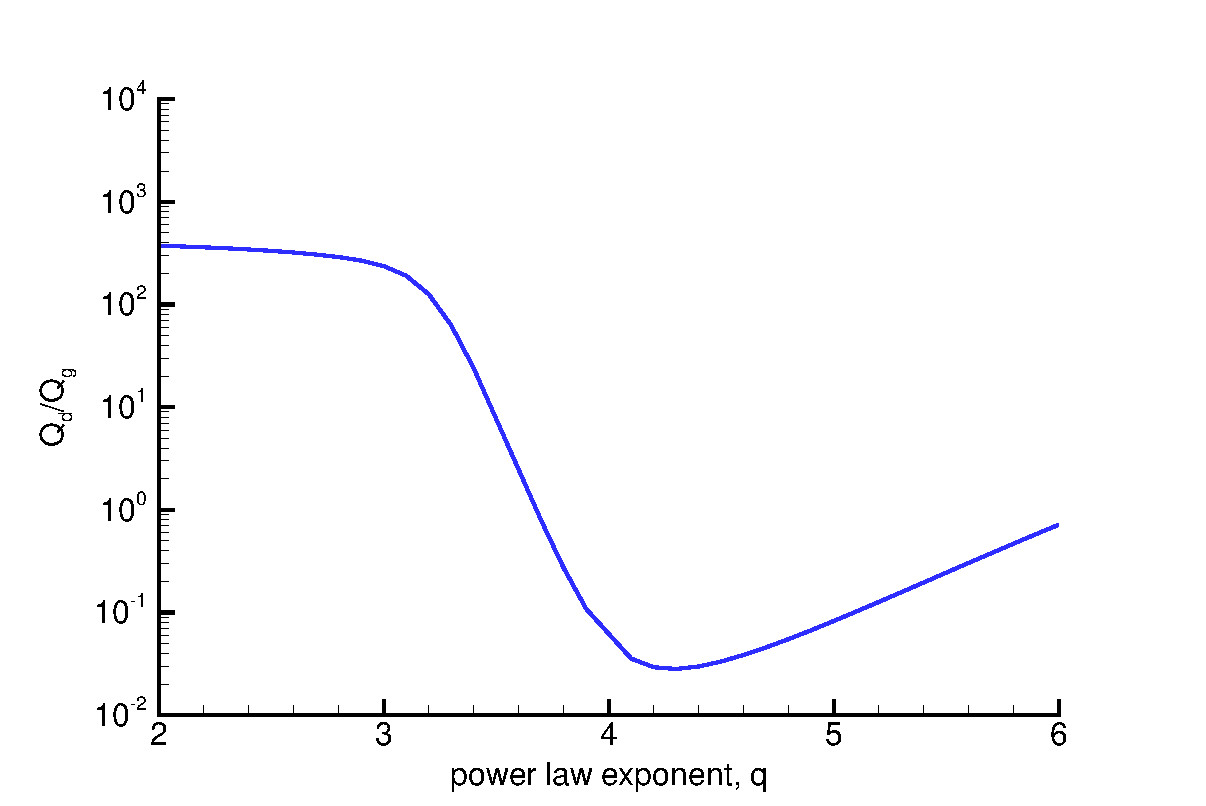}
 \caption{Dust-to-gas mass production rate ratio vs. power-law exponent for constant dust brightness.}			
 \label{fig_chi_q}
\end{figure}

For a given gas production rate $Q_g$ the maximum dust size $a_{max}$ is also constant. Fig.~\ref{fig_chi_q} shows how the dust-to-gas mass loss rate $Q_d/Q_g$ varies as a function or the power-law exponent of the dust size distribution for the same brightness. With increasing power-law exponent $q$ from minimal value to $\approx 3$ the $Q_d/Q_g$ is slowly decreasing (since in this case practically all dust mass is concentrating in a narrow range of largest sizes), but with increasing $q$ from 3 to 4 the ratio $Q_d/Q_g$ is strongly decreasing. For $4.5<q<6$ $Q_d/Q_g$ is increasing (within one order of magnitude). This inflection point of $Q_d/Q_g$ occurs at the transition from dust grains distribution with most mass being in the large dust sizes to where most mass is in the small dust sizes.

The growth of $Q_d/Q_g$ for $q>4.5$ is a consequence of the strong decrease of $q_{sca}$ for small sizes, therefore, in order to maintain the same brightness, it is necessary to eject more dust.

We should remember that this analytical result (Fig.~\ref{fig_chi_q}) assumed several
important simplifications:

\begin{enumerate}
    \item we assumed that the dust expansion is strictly radial;
    \item for evaluation of the dust brightness we used simplified optical properties (e.g. isotropic phase function);
    \item we assumed that gravity plays only a minor role;
    \item we assumed that the dust does not affect the gas flow. 
\end{enumerate}
We will discuss in the next section how these assumptions (in particular (1) and (3)) change the result.

\section{Results and Discussion}\label{sec:results}
\begin{figure}[h!]
	\includegraphics[width=\textwidth]{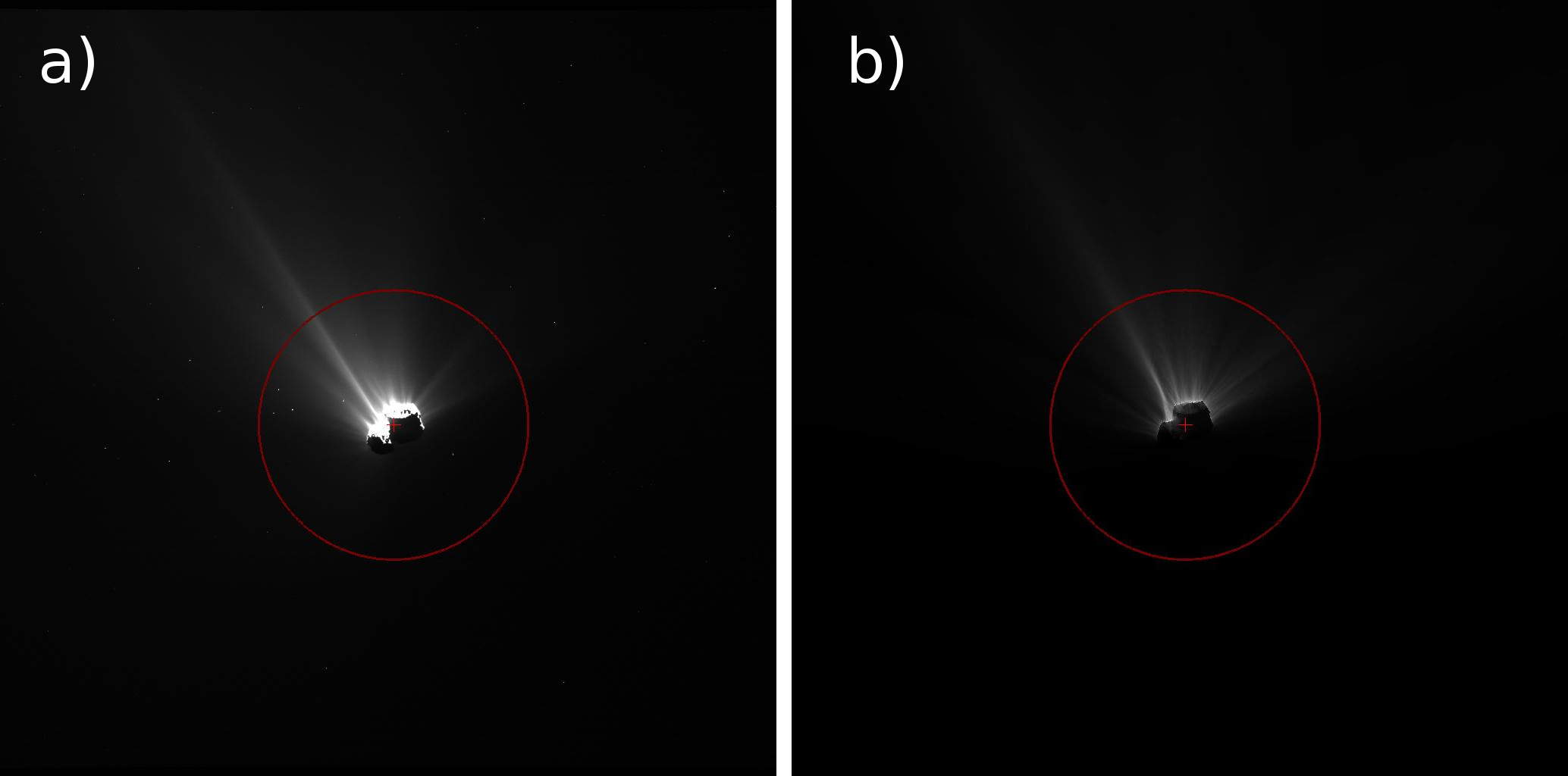}
	\caption{Panel a) shows the OSIRIS image WAC\textunderscore2015-08-09T09.13.16.574Z of epoch 12 (solstice) with an enhanced contrast to show the dust coma. Panel b) shows the synthetic image with power-law exponent of $q=4$ of our dust model. The crosses in each panel indicate the centre of the nucleus and the red circles indicate an impact parameter distance of $b=11$~km. }
	\label{fig:osiris-vs-sim} 
\end{figure}

To convolve the results of the dust dynamics model with the scattering properties to arrive at synthetic OSIRIS images we need to assume a dust size distribution. As is commonly done we presume that the number of particles, $n$, of a certain radius, $r$, follows a power-law:
\begin{equation}
    n(r) \propto r^{-q}\quad,
\end{equation}
where $q$ is the differential power-law exponent. Fig.~\ref{fig:osiris-vs-sim} illustrates an example of an OSIRIS and synthetic image for epoch 12 (solstice). As described in Sec.~\ref{sec:dust-dynamics}, we extract the integrated brightness along a circle with a constant impact parameter of $b=11$~km where possible (illustrated in the figure with the red circles). We should stress here again that it was not the aim of this work to match the emission distribution on the surface and thus all the structures in the coma.\\

For a given gas production rate, the three major factors controlling the brightness (see Eq.~(\ref{eqn_I}) for more detail) of the dust coma are: 
\begin{enumerate}
    \item the dust-to-gas mass production rate ratio, $Q_d/Q_g$, at the surface;
    \item the dust size distribution (i.e. the power-law exponent, $q$) at the surface;
    \item the scattering properties of the dust particles. 
\end{enumerate}
We should note that although we assume a uniform surface (i.e. globally constant $Q_d/Q_g$ and $q$) the actual values at each facet vary depending on the local gas flux. If a particular facets' local flux is too low to lift a certain particle size the resulting dust flux and size distribution from that surface facet will differ locally from the nominal values. The three input parameters above are not known a priori and are thus initially free parameters and in need of constraints. We have fixed the scattering properties by using published results that fit OSIRIS data \citep[see Sec.~\ref{sec:scattering} and][]{Markkanen2018}. This reduces the above parameters from three to two.\\

There are three further quantities that influence the coma brightness, but as we will show below their influence is small compared to the ones mentioned above. These are:
\begin{enumerate}
    \item the smallest dust size, $r_{min}$;
    \item the largest dust size, $r_{max}$;
    \item the bulk density of the dust, $\rho_d$.
\end{enumerate}
Of these three we will explore the influence of $r_{min}$ and $\rho_d$ on our results. We will not be artificially truncate the size range at large sizes by varying $r_{max}$. On the contrary, $r_{max}$ will be naturally regulated due to the balance of forces at the surface. If a given local gas flux is not sufficient for lifting a certain size, that will determine the largest dust size from that surface element.

Apart from the parameters that directly influence the brightness, several indirect factors further constrain the curves $Q_d/Q_g$ and $q$. We will discuss these constraints at the end of this section.\\

\begin{figure}[h!]
	\includegraphics[width=\textwidth]{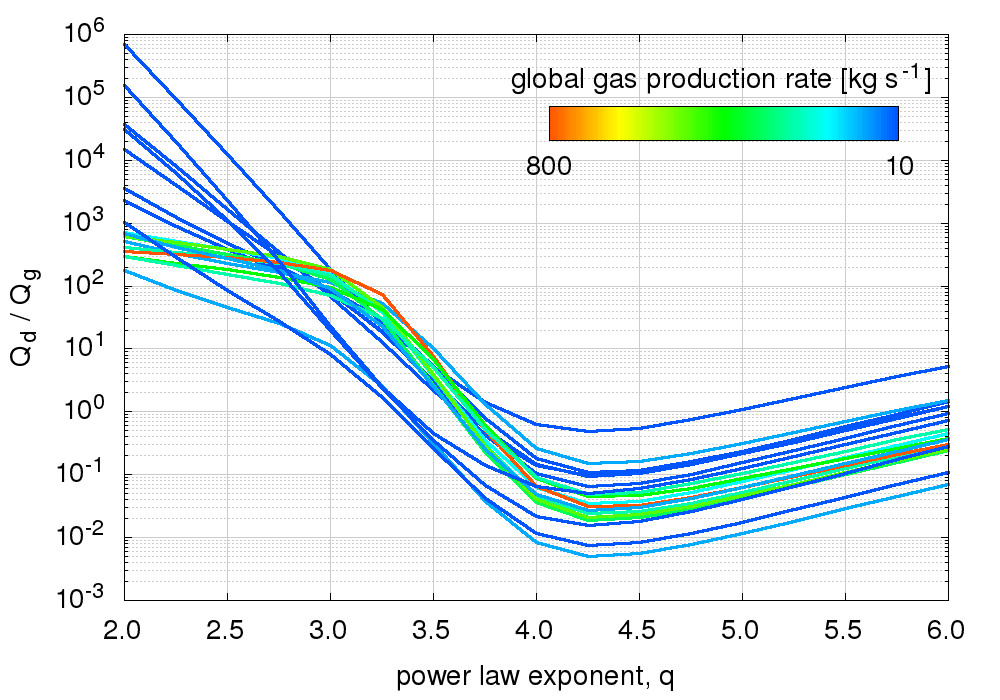}
	\caption{The dust-to-gas mass production rate ratio, $Q_d/Q_g$ as a function of the power-law exponent, $q$, is shown for each of the 18 OSIRIS images of this study. Each line represents an equal brightness curve where the respective $q$ and $Q_d/Q_g$ result in a synthetic image that matches the OSIRIS brightness. The colours of the curves indicate the global gas production rate of the respective epoch.}
	\label{fig:q-vs-D2G} 
\end{figure}

As has already been shown in Fig.~12 of \cite{Marschall2016} $Q_d/Q_g$ and $q$ are not independent. Knowing the brightness of the coma these two parameters constrain each other to a limiting set of parameter pairs. Fig.~\ref{fig:q-vs-D2G} shows $Q_d/Q_g$ as a function of $q$ for each of the 18 OSIRIS images of this study. Each line represents an equal brightness curve where the respective $q$ and $Q_d/Q_g$ result in a synthetic image that matches the OSIRIS brightness. Several things are noteworthy. First, all curves show minima between $q=4$ and $q=4.5$ and thus illustrate the inherent degeneracy between $Q_d/Q_g$ and $q$. Second, all cases with shallow power-laws ($q<3$) require very large $Q_d/Q_g$ of at least $10$ but in most cases around $100$. Third, steep power-laws ($4<q<6$) in all but one case require much less dust mass, i.e. $Q_d/Q_g \le 1$ to match the brightness of OSIRIS. Fourth, there is a clear trend in the gas production rate. As the gas production rate increases the slope in $Q_d/Q_g$ for shallow power-laws ($q<3$) becomes shallow, too. Or conversely for low gas production rates very high $Q_d/Q_g$ are needed to match the OSIRIS brightness when the power-law is shallow. This has to do with the amount of dust that can be lifted and escape the nucleus' gravity.\\

Comparing Fig.~\ref{fig:q-vs-D2G} to the analytical solution presented in Fig.~\ref{fig_chi_q} of Sec.~\ref{sec:analytic-solution} we see that for high gas production rates the model follows the analytical solution rather well. The places where we deviate from the analytical solution illustrate the effect of different physical processes. For the analytical solution we have assumed a  minor (but not negligible) role of gravity. The effect of gravity can be seen in the low gas production rate cases with shallow power-laws. There, in contrast to the analytical model which levels off at smaller power-law exponents, the dust coma model results in ever higher $Q_d/Q_g$. This is caused by the inability to lift large particles from the entire surface and therefore a higher $Q_d/Q_g$ is required to maintain the brightness. Thus the deviations at low gas production rates and shallow power-laws exhibit the non-minor role of the nucleus' gravity. As in the analytical model for steeper size distributions most mass is in the smallest particles, which are weakly scattering and thus hardly contribute to the brightness. This is compensated by an increase of $Q_d/Q_g$ at these steep power-laws.\\

Compared to previous work presented in Fig.~12 of \cite{Marschall2016} the $Q_d/Q_g$ values we find here (in particular for $q<3.5$) are much higher while the behaviour of the curves for steeper power-laws is within the expected range. The two main reasons we find larger values at shallow slopes are: 1) \cite{Marschall2016} assumed the scattering properties of astronomical silicate \citep{LaorDraine1993} which is much brighter than we now know; 2) we consider here considerably larger dust sizes as our upper limit. This extension of the size domain increases the $Q_d/Q_g$ by orders of magnitude because of high fall back fractions of dust that is gravitationally bound and weakly scattering.\\

\begin{figure}[h!]
	\includegraphics[width=\textwidth]{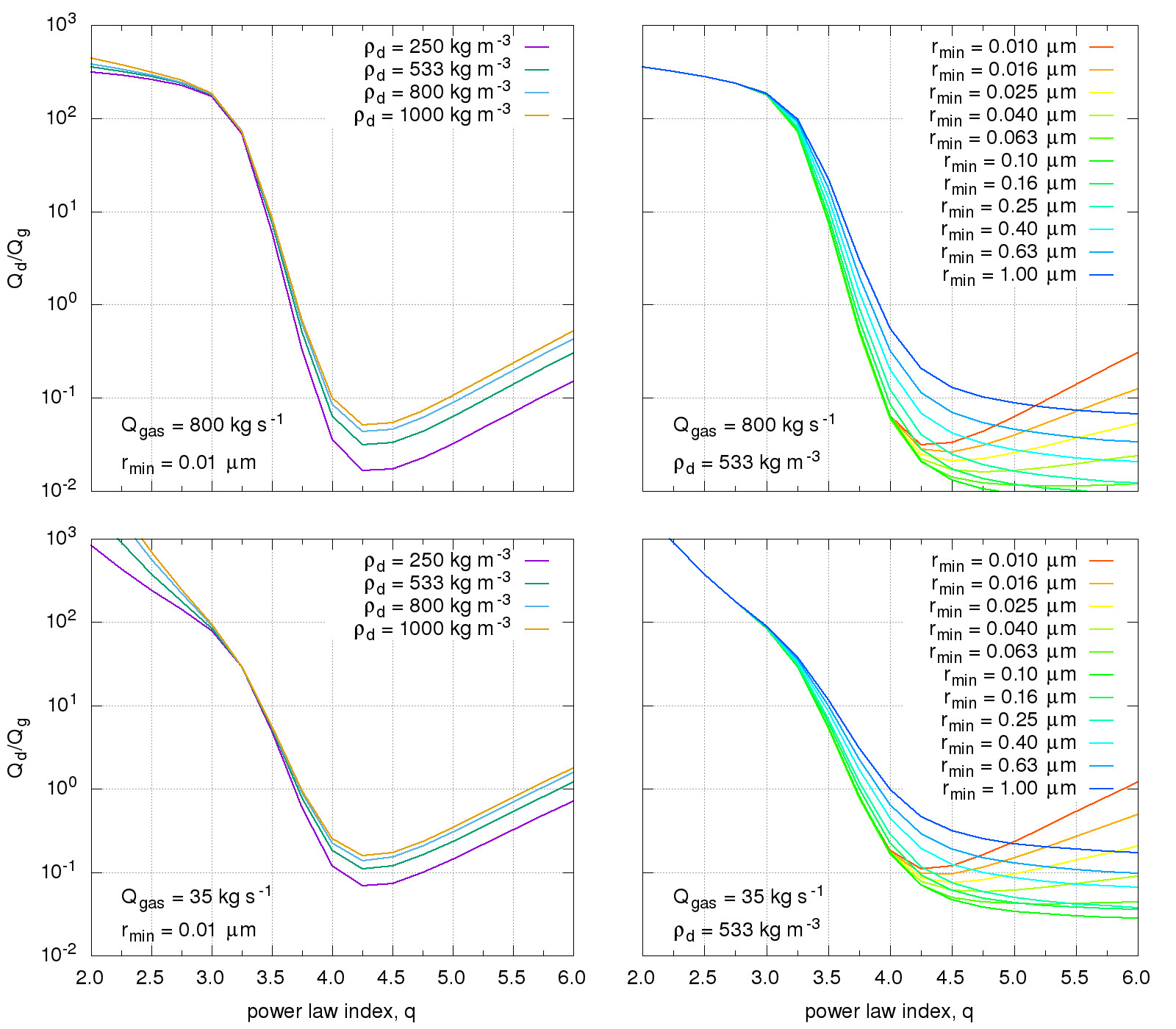}
	\caption{The four panels show the dust-to-gas ration as a function of power-law index. The left two panels show the results for different bulk dust densities ($250$~kg~m$^{-3}$ [purple], $533$~kg~m$^{-3}$ [green], $800$~kg~m$^{-3}$ [blue], $1000$~kg~m$^{-3}$ [orange]) assuming a minimum dust size of $0.01\mu$m and a gas production rate of $800$~kg~s$^{-1}$ (top panel, epoch 12) and $35$~kg~s$^{-1}$ (bottom panel, epoch 6). The two right panels show the results for varying minimum dust size of the power-law from $0.01\mu$m (red) to $1\mu$m (red) assuming a bulk dust densities of $533$~kg~m$^{-3}$ and a gas production rate of $800$~kg~s$^{-1}$ (top panel, epoch 12) and $35$~kg~s$^{-1}$ (bottom panel, epoch 6).}
	\label{fig:D2G-vs-varR-varRho} 
\end{figure}

Two assumptions going into Fig.~\ref{fig:q-vs-D2G} are worth discussing. First, we have assumed that all dust particles have a bulk density equal to the nucleus density ($533$~kg~m$^{-3}$). It is likely that the density of small particles is significantly larger than that and that the density then decreases with size. Because the exact relationship of the density as a function of dust size is currently unknown we have not tested a varying density as a function of size. But we have varied the bulk density for the entire range of dust sizes between $250$~kg~m$^{-3}$ and $1000$~kg~m$^{-3}$. The two left panels of Fig.~\ref{fig:D2G-vs-varR-varRho} show the results for a moderate activity environment (epoch 6 - inbound equinox, $Q_g=35$~kg~s$^{-1}$) and a high activity environment (epoch 12 - solstice, $Q_g=800$~kg~s$^{-1}$). For $3<q<3.75$ the differences between the different dust densities is minimal. For $q>3.75$ the differences are larger, in particular in the high activity case. How the bulk dust density impacts the total dust mass loss will be explored later in this section.

Second, we have currently assumed that the smallest dust size is $0.01\mu$m. This might not be the preferred choice and a much larger smallest size should be considered. The MIDAS instrument detected $1\mu$m particles \citep[e.g][]{Mannel2019} and there is indirect evidence of sub-micron particles observed by VIRTIS during outbursts \citep{BockeleeMorvan2017}. We have thus explored the range of the smallest sizes between $0.01\mu$m and $1\mu$m. The two right panels of Fig.~\ref{fig:D2G-vs-varR-varRho} explore the effect of the smallest size on the dust-to-gas ratio by varying the smallest size. Compared to the differences seen for different bulk dust density the effect of the smallest size is quite substantial. As we would expect the smallest size does not affect the result for $q<3$ as in these cases most of the mass is in the large particles. As $q$ increases from $3$ the curves for different smallest sizes start diverging. Two trends can be observed. As the smallest size increases from $0.01\mu$m to $\sim0.1\mu$m the dust-to-gas ratio starts to flatten out beyond $q=4$. This is caused by the fact that the size distribution is no longer dominated by very inefficiently scattering particles. As the smallest size continues to increase to $1\mu$m the overall dust-to-gas ratio increases. This is because the most efficient scatterers (see Fig.~\ref{fig_qfit2}) are being removed from the size distribution and must be compensated by more mass of all other sizes. For very steep power-laws the difference in the dust-to-gas ratio can be up to 1.5 orders of magnitudes. How the smallest size impacts the total dust mass loss will be explored later in this section.

\begin{figure}[h!]
	\includegraphics[width=\textwidth]{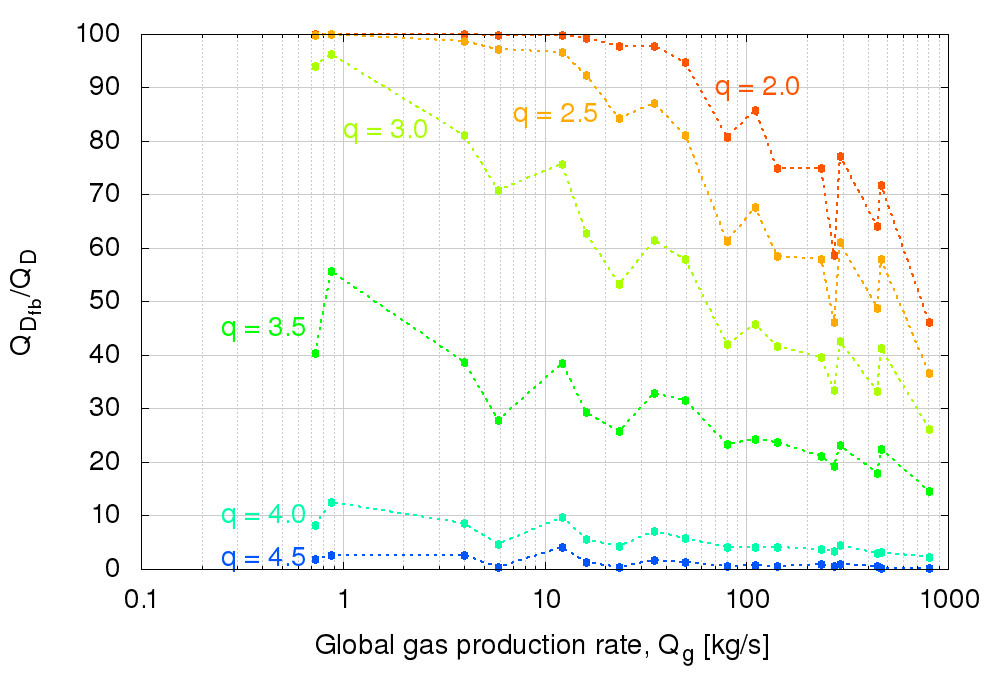}
	\caption{The ratio of dust mass falling back, $Q_{D_{fb}}$ to total dust mass, $Q_D$ is shown as a function of global gas production rate for different power-law exponents.}
	\label{fig:fallBackFraction} 
\end{figure}

How the fraction of gravitationally bound dust mass, which falls back to the nucleus, varies as a function of global gas production rate is shown in Fig.~\ref{fig:fallBackFraction} for different power-law exponents. This illustrates that for very low gas production rates and very shallow power-laws ($q<2.5$) almost the entire dust mass emitted from the surface will be redeposited. This explains the large increase seen in $Q_d/Q_g$ of Fig.~\ref{fig:q-vs-D2G}. Conversely, in the case of steep power-laws ($q\ge4.5$) almost all of the dust escapes the nucleus' gravity field irrespective of the gas production rate. In all cases, the fraction of fall back decreases as the gas production rate increases. Therefore, the fraction of fall back material decreases as the comet approaches the Sun. At large heliocentric distances, large fractions of dust emitted will return to the surface (i.e. $>50\%$ for $q<3.5$). But the gas and dust production rates are highest at perihelion/solstice, thus the total amount of fall back during one apparition is dominated by the fraction of fall back during that period.\\

\begin{figure}[h!]
	\includegraphics[width=\textwidth]{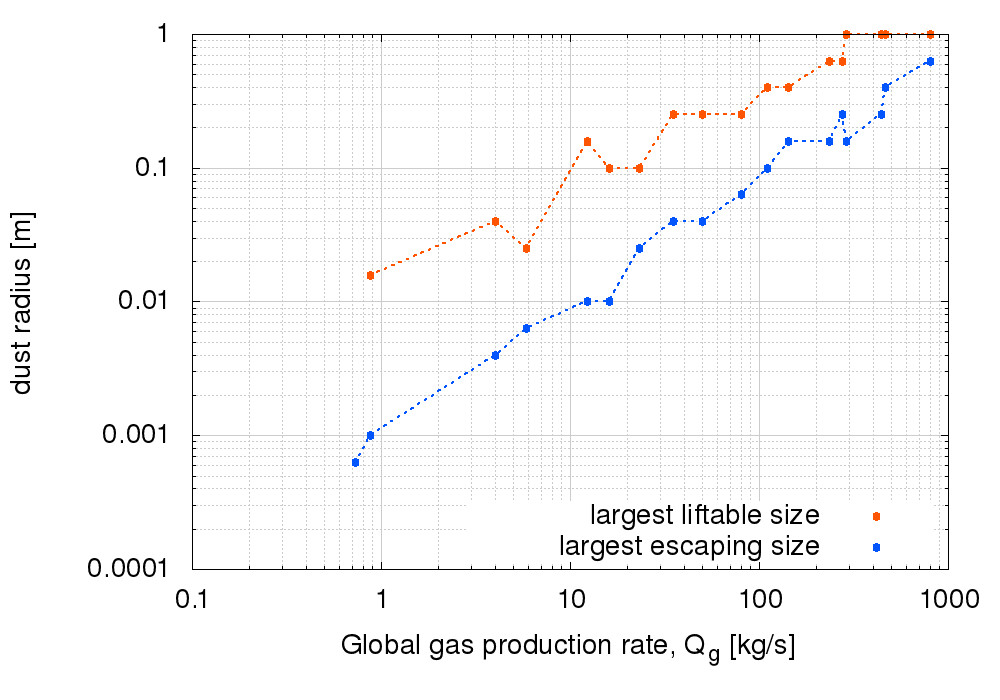}
	\caption{The largest liftable dust size (red curve) and the largest escaping dust size (blue curve) are shown as a function of the gas production rate.}
	\label{fig:largestSize} 
\end{figure}

The fraction of fall back is also tightly bound to the maximum liftable dust size. Fig.~\ref{fig:largestSize} shows as a function of global gas production rate the largest dust size that can still be lifted from the surface of the nucleus. The figure also shows the largest dust size that can escape the gravity field of the comet. As the gas production rate increases so do the largest liftable and escaping dust sizes. For $Q_g>300$~kg~s$^{-1}$ the largest liftable dust size is larger or equal to 1~m, which is the largest size in our model. Though these sizes, or larger, can be lifted they will not be able to escape the gravity field of the comet and will be redeposited on the surface. The largest size that can escape the comet at peak gas production is roughly $0.6$~m. We should note though that this calculation neglects surface cohesion, solar radiation pressure, and heat transport to the subsurface that is needed to eject such large particles. Here we only consider the balance of gas drag and gravity to determine these largest sizes.\\

The discussion of the previous paragraphs illustrates that multiple properties of the dynamical simulation of the dust coma (size distribution, dust-to-gas ratio, and the fraction of fallback) as well as the optical properties of the dust are not independent but mutually constraining. E.g. a given fraction of fallback implies a certain size distribution which in turn constrains the possible dust-to-gas ratios for a particular set of scattering properties. Though for a particular single OSIRIS image these parameters can be constrained there still remains a rather large set of parameters that are consistent with the data as presented to this point (including dust coma brightness in the OSIRIS images and local gas densities of ROSINA/COPS). \\

\begin{figure}[h!]
	\includegraphics[width=\textwidth]{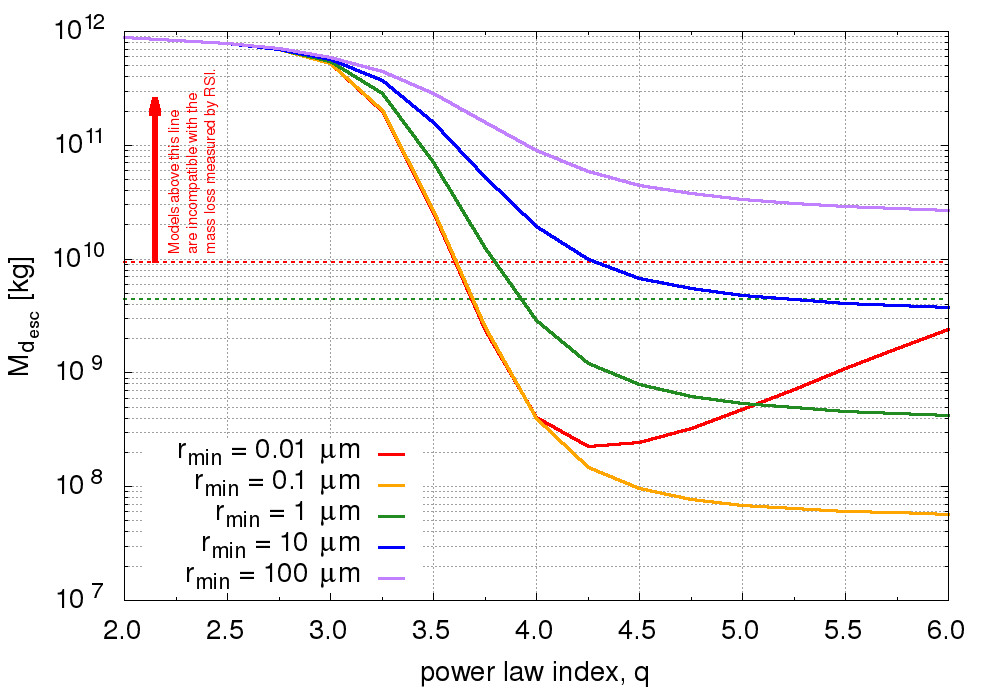}
	\caption{The total escaping dust mass, $M_{d}^{esc}$, is shown as a function of the power-law index for five different minimum dust radii, $r_{min}$. The horizontal dashed lines show the nominal ejected dust mass (green) and the maximum ejected dust mass (red).}
	\label{fig:massLoss-vs-q} 
\end{figure}

While we have only considered constraints within each epoch there is one strong constraint covering the entire mission. That is the measurement of the total mass loss during the Rosetta apparition. During the 2 year mission comet 67P had lost $(10.5\pm3.4)\cdot10^9~$kg \citep{Paetzold2019}. 
The total mass loss, $M_{tot} = (10.5\pm3.4)\cdot10^9~$kg,  is:
\begin{equation}
    M_{tot} = M_{g} + M_{d}^{esc} \quad ,
\end{equation}
where $M_{g}$ is the total volatile mass loss, and $M_{d}^{esc}$ is the total escaping dust mass. For the dust masses, we can further specify that
\begin{equation}
    M_{d} = M_{d}^{esc} + M_{d}^{fb} \quad ,
\end{equation}
where $M_{d}$ is the total dust mass ejected from the nucleus, and $M_{d}^{fb}$ is the dust mass that falls back to the surface. We have determined the total volatile mass loss to be $(6.1 \pm 1.5) \cdot 10^9$~kg. Combined with the total mass loss of the nucleus it follows that $M_{d}^{esc} = (4.4^{+4.9}_{-4.4}) \cdot 10^9$~kg. Note that within this interval exists the possibility that $M_{d}^{esc} = 0$~kg. Though we know that dust escaped from the nucleus the simple mass balance would not exclude this possibility. We can now integrate the total dust mass loss over the orbit of the comet for different power-law exponents. For the integration, we assume a linear interpolation of the results between epochs. Figure~\ref{fig:massLoss-vs-q} shows the $M_{d}^{esc}$ as a function of the power-law exponent, $q$. Cases, where $M_{d}^{esc}$ exceeds the nominal dust mass loss of $4.4 \cdot 10^9$~kg (horizontal dashed green line) or the maximum dust mass loss of $9.3 \cdot 10^9$~kg (horizontal dashed red line), can be discarded. Furthermore, where the mass loss curve intersects the mass loss indicates the corresponding power-law that fits the data. Figure~\ref{fig:massLoss-vs-q} also illustrates the effect of the smallest size - discussed earlier in more detail for an individual OSIRIS image - on the total mass loss. The effect of the smallest size is rather limited for $0.01 \mu$m~$ < r_{min} < 1\mu$m because the dust mass curves cross between $3.5<q<4$. This also implies that the effect of the bulk dust density is even smaller than the effect from the choice of the smallest size (see discussion about Fig.~\ref{fig:D2G-vs-varR-varRho} above). We can also see that for $r_{min} > \sim 12\mu$m there will no longer be a nominal solution to the constraints. Further, for $r_{min} > \sim 30\mu$m there is no solution at all because the curve will stay above the maximum escaping mass for all power-law exponents studied here. This means that the minimum dust size must be strictly small than $\sim30\mu$m and nominally even smaller than $\sim12\mu$m. Fig.~\ref{fig:massLoss-vs-q} illustrates how we can determine the power-law exponent for the nominal and maximum dust mass loss, which in turn determines the dust-to-gas ratio, dust production rates, fraction of dust fallback. As the minimum size grows larger than $1\mu$m the required power-law exponent increases and becomes rather large. There remains the issue of the minimum escaping mass. As discussed above the lower limit according to the total and volatile mass loss is zero. But for our models, the minimum escaping dust mass is never zero. We have thus chosen the smallest possible mass loss of each model as the minimum mass loss. The resulting power-law exponents, dust mass losses, dust-to-gas ratios, and fall back fraction are summarised in Tab.~\ref{tab:result}. We have also determined the deposition height, $H$, that results if the fallback material is spread equally on the smooth deposits (9.43~km$^2$) identified by \cite{Thomas2018}.

\begin{table}[]
\caption{Power-law exponents, dust mass loss, dust-to-gas ratio, and fall back as a function of the smallest dust size (see Fig.~\ref{fig:massLoss-vs-q}) . Dashed entries mean that no solution is possible for this size.}
\begin{tabular}{r|lllll}\label{tab:result}
$r_{min}$ &   q   &  $M_{d}$ [kg] &  $M_{d}^{esc}$ [kg] &  $M_{d}^{fb}$ [kg] \\ \hline\hline
$0.01\mu$m  & $3.7^{+0.57}_{-0.078}$ & $5.1^{+6.0}_{-4.9}\cdot10^{9}$ &  $4.4^{+4.9}_{-4.2}\cdot10^{9}$ &  $6.8^{+11}_{-6.8}\cdot10^{8}$\\
$0.1\mu$m  & $3.7^{+2.3}_{-0.079}$ & $5.1^{+6.0}_{-5.0}\cdot10^{9}$ &  $4.4^{+4.9}_{-4.4}\cdot10^{9}$ &  $6.7^{+11}_{-6.7}\cdot10^{8}$\\
$1\mu$m  & $3.9^{+2.1}_{-0.13}$ & $4.7^{+5.6}_{-4.3}\cdot10^{9}$ &  $4.4^{+4.9}_{-4.0}\cdot10^{9}$ &  $3.1^{+7.3}_{-3.1}\cdot10^{8}$\\ 
$10\mu$m  & $5.2^{+0.80}_{-0.93}$ & $4.5^{+5.1}_{-0.69}\cdot10^{9}$ &  $4.4^{+4.9}_{-0.66}\cdot10^{9}$ &  $0.35^{+1.6}_{-0.35}\cdot10^{8}$\\
$100\mu$m  & - & - & - & -\\\\\hline\hline

$r_{min}$  & $M_{d}$/$M_{g}$ & $M_{d}^{esc}$/$M_{g}$ & $M_{d}^{fb}$/$M_{d}$ [$\%$]  &  $H$ [cm] \\\hline\hline
$0.01\mu$m  & $0.84^{+1.6}_{-0.81}$ & $0.73^{+1.3}_{-0.70}$ & $13^{+2.6}_{-12}$ & $14^{+22}_{-14}$\\
$0.1\mu$m  & $0.84^{+1.6}_{-0.83}$ & $0.73^{+1.3}_{-0.72}$ & $13^{+2.6}_{-12}$ & $13^{+21}_{-13}$\\
$1\mu$m  & $0.78^{+1.5}_{-0.73}$ & $0.73^{+1.3}_{-0.67}$ & $6.6^{+3.4}_{-5.9}$ & $6.3^{+14}_{-6.2}$ \\
$10\mu$m  & $0.74^{+1.4}_{-0.24}$ & $0.73^{+1.3}_{-0.23}$ & $0.78^{+1.3}_{-0.78}$ & $0.69^{+3.2}_{-0.69}$ \\
$100\mu$m  & - & - & - & - 
\end{tabular}
\end{table}

The results in Tab.~\ref{tab:result} show that the integrated quantities are rather insensitive to the choice of the smallest dust size if $r_{min}\le 1\mu$m. For minimum sizes larger than $1\mu$m the power-law becomes steeper and thus the amount of dust fall-back goes down. The dust-to-gas ratio is rather stable for all cases and is of the order of $0.8$ with an error of the order of $100\%$. This means that while the nominal case reflects a comet that contains more volatiles than dust the case of a dusty comet lies within the error.

The fallback in all cases is of the order of $10\%$ and results in a deposition height of the order of $10$~cm. Because the deposition is likely non-uniform it is therefore easily thinkable that in certain areas dust of the order of metres is deposited while in others only a few centimetres.

This analysis assumes that the dust size distribution does not change along the orbit. There is an indication \citep[e.g.][]{Merouane2018} that this is not the case and that the slope is varying with heliocentric distance. Our model cannot resolve/constrain this. All the quantities here are heavily dominated by the period around perihelion and summer solstice when the emission was the highest. Therefore, the power indexes found here reflect the values for this period.

The power-law we find to be compatible with the data is an independent result based only on the brightness of the dust coma and the total mass loss balance. Because most mass is ejected around perihelion, this power-law mainly reflects this period and deviations of it at larger heliocentric distances \citep{Fulle2016} would not influence the result. The value we find is in line with other measurements around perihelion e.g. the in-situ measurement of $q=3.7$ by GIADA \citep{Fulle2016}, $q=3.8$ by COSIMA \citep{Merouane2018}, as well as ground-based estimates for the dust tail of $3.6 <q <4.3$ for sizes smaller than 1~mm and $q=3.6$ for sizes larger than 1~mm \citep{Moreno2017}.

A check of our dust dynamics model is the comparison of our model dust speeds with the ones measured by GIADA. For the period between 2.2~AU inbound to 2.0~AU outbound \citet{DellaCorte2016} have reported 141 dust particle detection for which a dust speed and mass could be determined. Of these, the smallest particle had a mass of $2.8\cdot10^{-9}$~kg, which corresponds to a radius of $108~\mu$m assuming a spherical particle with our nominal dust density. The measured dust speeds varied between $0.3$ and $34.8$~m~s$^{-1}$ \citep{DellaCorte2016}. A further constraint is the fact that Rosetta spent $\sim 65\%$ of the time at phase angles of $\sim 90^\circ$ and an additional $\sim 25\%$ of the time at phase angles of $\sim 60^\circ$ which implies that the particles were mainly collected in those locations \citep[see also Fig.~5 in][]{DellaCorte2016}. At a phase angle of $90^\circ$ our model dust particles with radius $100~\mu$m have speeds of $(3.5\pm1.0)$~m~s$^{-1}$ at the inbound equinox (epoch 6) and $(18\pm5.6)$~m~s$^{-1}$ at the summer solstice (epoch 12). For phase angles of $60^\circ$ the model dust particles have speeds of $(7.0\pm1.2)$~m~s$^{-1}$ at the inbound equinox and $(32\pm5.2)$~m~s$^{-1}$ at the summer solstice. Our dust speeds are thus well in line with the measured speeds by GIADA given that larger particles will have lower speeds than the ones listed above.

We should highlight that our peak dust production rate ($\sim530$~kg~s$^{-1}$) is roughly an order of magnitude lower than those reported by e.g. \cite{Moreno2017} ($\sim3000$~kg~s$^{-1}$) or \cite{Ott2017} ($\sim 8300$~kg~s$^{-1}$). Furthermore, \cite{Moreno2017} report a total dust mass loss of $1.4\cdot0^{10}$~kg. As neither \cite{Moreno2017} nor \cite{Ott2017} report any error bars on their results, we cannot asses if they are plausible. If taken at face value both results are inconsistent with the RSI measurement of the total mass loss of the comet \citep{Paetzold2019} taking into account the estimates of the volatile mass loss in this work and others \citep{Combi2020,Laeuter2018}.

Finally, the determination of the power-law exponent allows us to determine the dust production rate (Fig.~\ref{fig:Q-vs-epoch}, top panel, green band), dust-to-gas ratio (Fig.~\ref{fig:Q-vs-epoch}, bottom panel, green line), and fraction of fallback (Fig.~\ref{fig:Q-vs-epoch}, bottom panel, purple line) as a function of time. The dust production rates are linearly interpolated between the epochs. Unfortunately, our model is rather noisy but the overall trends are robust enough that we feel comfortable making further conclusions. The fraction of dust fallback is highest at large heliocentric distances and then decreases towards perihelion and reaches its minimum at summer solstice where the activity peaked. Though the faction of fallback is smallest at the peak of the activity (solstice), most mass that is falling back will still be from the period of summer solstice because of the high activity. The behaviour of the fraction of dust fallback is symmetric for the inbound and outbound part of the comets' orbit. Contrary to that the dust-to-gas ratio is highest ($\sim1.5$) at large heliocentric distances inbound and keeps decreasing along the entire orbit and does not significantly increase on the outbound leg but rather flattens out at $\sim0.1$. This might be indicative of the comet shedding its dust mantle, in particular in the northern hemisphere. This trend of decreasing dust-to-gas ratio along the orbit manifests itself also in the asymmetry of the global dust production. To first order, the dust production rate follows the gas production rate during the inbound leg but the dust production rate drops faster than the gas production rate post solstice. This is also observed in ground-based measurements \citep{Moreno2017}. There is also an intriguing spike in the dust-to-gas ratio after the inbound equinox coinciding with an increase in the total dust production rate. Future in-depth work will be needed to confirm the nature of this feature which does not seem to be present in the observations of the outer coma from ground-based measurements. But if it is truly there it can be understood as the comet shedding its southern dust mantle because the feature coincides with the period when the southern hemisphere receives increasing insolation.

\section{Summary and Conclusions}\label{sec:summary-conclusion}
In this work, we have simulated the inner gas and dust coma of comet 67P covering the entire Rosetta mission by splitting it into 20 epochs. The gas production rates of each epoch were constrained by in-situ measurements of the gas density by ROSINA/COPS. From that, the total gas mass loss is estimated at $(6.1 \pm 1.5) \cdot 10^9$~kg. This is in line with values published in other works as e.g. $(6.3 \pm 2.0) \cdot 10^9$~kg \citep{Combi2020} or $(5.8 \pm 1.8) \cdot 10^9$~kg \citep{Laeuter2018}. It also illustrates that it is not necessary to know the surface-emission distribution well to estimate the total global volatile loss.\\

By simulating synthetic OSIRIS images of the dust coma we showed how the dynamical and optical properties of the dust can be constrained. In particular, we showed how the dust-to-gas mass production rate ratio, $Q_d/Q_g$, the power-law exponent, $q$, the fraction of dust fall back, $Q_{d}^{fb}$, and the scattering properties are inter-related and constrain each other. Because these parameters are not independent they need to be fit simultaneously. E.g. the lowest mass needed to match the brightness of the dust coma as observed by OSIRIS is achieved with power-law distributions with exponents between 4 and 4.5. Using the constraint of the total mass loss of the comet during the 2015 apparition we were able to show that only a narrow parameter set fits all observations. 
We determined that power-laws with $q=3.7^{+0.57}_{-0.078}$ are consistent with the data. This results in a total of $5.1^{+6.0}_{-4.9}\cdot10^9$~kg of dust being ejected from the nucleus surface, of which $4.4^{+4.9}_{-4.2}\cdot10^9$~kg escape to space and $6.8^{+11}_{-6.8}\cdot10^8$~kg (or an equivalent of $14^{+22}_{-14}$~cm over the smooth regions) is re-deposited on the surface. This leads to a dust-to-gas ratio of $0.73^{+1.3}_{-0.70}$ for the escaping material and $0.84^{+1.6}_{-0.81}$ for the ejected material. Further, the minimum dust size must be strictly smaller than $\sim30\mu$m and nominally even smaller than $\sim12\mu$m. We have found that these results are robust with respect to varying the smallest dust size between $0.01-1\mu$m and variations in the bulk density of the dust between $250-1000$~kg~m~$^{-3}$. \\

It remains an open question as to how dust particles are lifted/ejected from cometary surfaces \citep[see e.g.][]{Vincent2019}. Furthermore, a more detailed study of the change in the dust size distribution with heliocentric distance would be of great interest and could refine the work presented here. Finally, comprehensive work on estimating the amount of dust deposition through e.g. local digital terrain modelling \citep[e.g. method by][]{Jorda2016} would provide valuable additional constraints.

\section*{Author Contributions}
RM performed the modelling of the gas and dust coma as well as the comparisons with ROSINA/COPS and OSIRIS and the analysis related to that. JM performed the calculations of the scattering properties of the dust particles. 
SBG implemented new features that optimise \texttt{DRAG3D}. OPR produced the unstructured simulation grid within which all \texttt{UltraSPARTS} and \texttt{DRAG3D} were run. NT wrote the IDL programs for reading and analysis of the OSIRIS images. JW provided support for running and optimisation of \texttt{UltraSPARTS}.

\section*{Funding}
Raphael Marschall acknowledges the support from the Swiss National Science Foundation grant 184482.\\
Johannes Markkanen acknowledges the support from ERC Grant No. 757390.\\
The team from the University of Bern is supported through the Swiss National Science Foundation, and through the NCCR PlanetS.

\section*{Acknowledgments}
We thank Frank Preusker and Frank Scholten for providing us the comet shape model SHAP7 \cite{Preusker2017} used in this work.\\
We thank Vladimir Zakharov for providing valuable comments on the section of the analytical solution. \\ 
We acknowledge the personnel at ESA's European Space Operations Center (ESOC) in Darmstadt, Germany, European Space Astronomy Center (ESAC) in Spain, and at ESA for the making the Rosetta mission possible. Furthermore, we thank the OSIRS and ROSINA instrument and science teams for their hard work. We thank Martin Rubin and Kathrin Altwegg for giving us access and support to/for the ROSINA/COPS data.



\bibliographystyle{frontiersinSCNS_ENG_HUMS} 
\bibliography{frontiers-4arxiv}


\begin{table}[h!]
\caption{The mean gas production rate, $\bar q_g$ [kg s$^{-1}$] as a function of ephemeris time (ET): \\$\bar q_g(ET) = a\cdot {ET}^2 + b\cdot ET + c$\\}
\begin{tabular}{r|rrr}\label{tab:fitting-parameters-q}
epoch & a    & b        & c       \\ \hline\hline
2     & 2.400340e-14   & -2.204283e-05  &  5.061404e+03 \\
3     & 9.908836e-14   & -9.285540e-05  &  2.175673e+04 \\
4     & 3.151072e-13   & -2.985843e-04  &  7.073810e+04 \\
5     & 2.843195e-13   & -2.690371e-04  &  6.364899e+04 \\
6     & 4.705372e-12   & -4.537366e-03  &  1.093862e+06 \\
7     & 4.739402e-12   & -4.570374e-03  &  1.101867e+06 \\
8     & 2.736443e-12   & -2.620080e-03  &  6.271145e+05 \\
9     & -8.310099e-14  &  1.344634e-04  & -4.564429e+04 \\
10    & 8.356934e-12   & -8.136523e-03  &  1.980682e+06 \\
11    & 2.648205e-11   & -2.595811e-02  &  6.361453e+06 \\
12    & -9.326697e-11  &  9.223562e-02  & -2.280309e+07 \\
13    & 2.625803e-11   & -2.622753e-02  &  6.549455e+06 \\
14    & 1.025950e-11   & -1.029856e-02  &  2.584549e+06\\
15    & 2.918884e-12   & -2.954660e-03  &  7.477641e+05\\
16    & 1.108671e-12   & -1.134214e-03  &  2.900836e+05\\
17    & 1.322669e-12   & -1.350748e-03  &  3.448580e+05\\
18    & 1.589655e-13   & -1.643041e-04  &  4.245606e+04\\
19    & 5.313567e-14   & -5.537983e-05  &  1.442952e+04\\
\end{tabular}
\end{table}





\end{document}